\documentclass[fleqn,12pt,twoside]{article}
\usepackage{espcrc1}

\usepackage{graphicx}
\usepackage[figuresright]{rotating}

\newcommand{\AmS}{{\protect\the\textfont2
  A\kern-.1667em\lower.5ex\hbox{M}\kern-.125emS}}
\newcommand{\beq}{\begin{equation}}
\newcommand{\eeq}{\end{equation}}
\def\al{\alpha}

\def\tr{{\mbox{Tr}}}
\def\re{{\Re\mbox{e}}}
\def\im{{\Im\mbox{m}}}

\title{Correlations for non-Hermitian Dirac operators: chemical potential in 
three-dimensional QCD}

\author{G. Akemann\address{
Service de Physique Th\'eorique$^\dagger$, CEA/Saclay, 
F-91191 Gif-sur-Yvette Cedex, France
}%
        \thanks{Work supported in part by EU TMR grant 
ERBFMRXCT97-0122.\newline 
$^\dagger$Unit\'e associ\'e CNRS/SPM/URA 2306
}}
       
\begin{document}

\maketitle

\begin{abstract}
In the presence of a non-vanishing chemical potential the
eigenvalues of the Dirac operator become complex. We 
use a Random Matrix Model (RMM) approach to calculate analytically 
all correlation functions at weak and strong non-Hermiticity for 
three-dimensional QCD with broken flavor symmetry and 
four-dimensional QCD in the bulk.
\end{abstract}

\section{The Model}

In \cite{Steph} a RMM has been introduced
to study the influence of a chemical potential on the support of 
Dirac operator eigenvalues. Here, we aim to derive 
microscopic correlations of complex eigenvalues at zero virtuality 
which are important for spontaneous symmetry breaking. 
For simplicity we consider a non-chiral RMM relevant 
for three-dimensional QCD with broken flavor symmetry \cite{VZ} 
as well as in four dimension away from zero. 
In the spirit of \cite{VZ} we replace the QCD Dirac operator with a 
constant complex matrix of size $N\times N$ 
as the Dirac eigenvalues become complex at finite density: 
\beq
{\cal Z}_{QCD3}^{(2N_f)}(\{m_f\}) \ =\ \int dJdJ^{\dagger}
\prod_{f=1}^{N_f}\left|\det[J-im_f]\right|^2 
 \exp\left[-\frac{N}{1-\tau^2}\tr\left(JJ^\dagger-\tau\re
     J^2\right)\right] \ .
\label{Z}
\eeq
The measure follows from taking the same 
Gaussian distribution for both the Hermitian and anti-Hermitian parts 
of $J$, $H$ and $i[(1-\tau)/(1+\tau)]^{1/2}A$, respectively.
We present results in the limit of weak non-Hermiticity where 
$\lim_{N\to\infty}2N(1-\tau)\equiv\alpha$ is kept fixed as well as in the 
limit of strong non-Hermiticity with $\tau\in[0,1)$.
In the weak limit the parameter $\alpha$ mimics the influence of the 
chemical potential.
The model (\ref{Z}) has been solved for $N_f=0$ both 
in the strong \cite{Gin} and weak limit  \cite{FKS}, where in the latter  
the existence of orthogonal (Hermite) polynomials in the 
complex plane was exploited. 
Here, we extend both results to include $2N_f$ fermions flavors,
massive or massless, using the same method as in \cite{AK}. 
For that reason we have to take the absolute value of the fermion determinant.

\begin{figure}[htb]
\centerline{
\includegraphics[width=19pc]{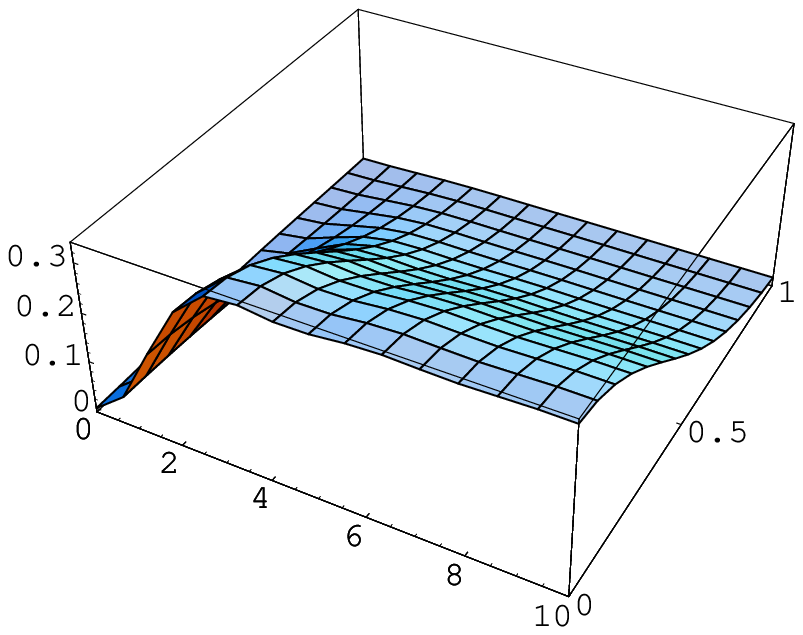}
\put(-160,20){$\re\ \xi$}
\put(-30,50){$\im\ \xi$}
\put(-220,170){$\rho_s^{(2)}(\xi)$}
\put(80,20){$\re\ \xi$}
\put(200,50){$\im\ \xi$}
\put(0,170){$\rho_s^{(2)}(\xi)$}
\includegraphics[width=19pc]{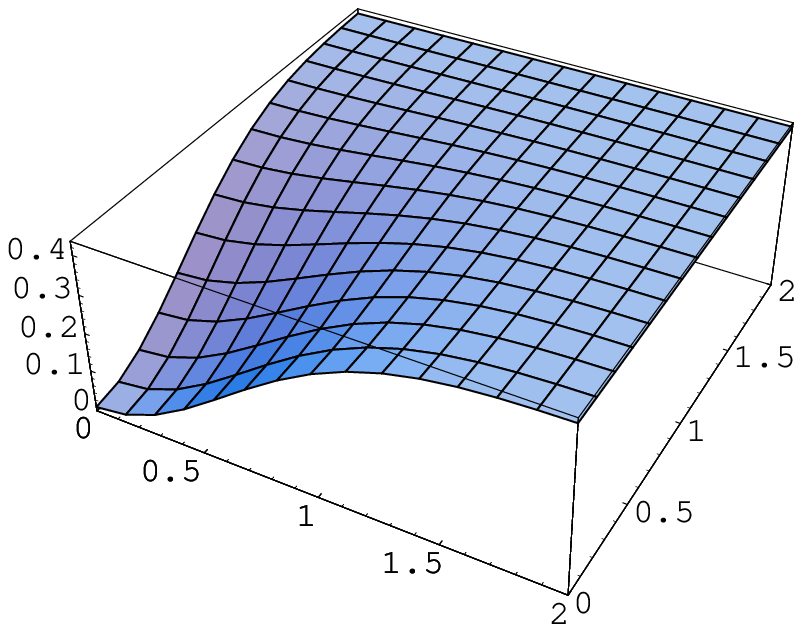}
}
\caption{
The 2-flavor massless density 
at weak (left) and strong non-Hermiticity (right).}
\label{rhow+s}
\end{figure}

\section{Results}

We present the two-flavor case as 
an example and refer to \cite{A01} for the general case. 
In the weak limit the 
eigenvalues $z$ of the matrix $J$ and the quark masses $m_f$ 
are rescaled as $Nz=\xi$ and $Nm_f=\mu_f$, respectively.  
The microscopic density shown 
in Fig. \ref{rhow+s} 
reads \cite{A01} 
\beq
\rho^{(2)}_S(\xi) \ =\ \frac{1}{\pi\al}
\exp\left[-\frac{2}{\al^2}\im^2\xi\right]\left\{ g_\al(2i\im\ \xi)
-\frac{g_\al(\xi+i\mu)g_\al(i\mu-\xi^\ast)}{g_\al(2i\mu)}\right\} \ ,
\label{rho1}
\eeq
with
$g_{\al}(\xi)\equiv\int_{-\Sigma}^\Sigma
du\exp\left[-\al^2u^2/2+i\xi u\right]/\sqrt{2\pi}$ 
and $\Sigma$ being the condensate.
It displays the exponential suppression in the complex plane as for $N_f=0$ 
\cite{FKS} and the oscillations and level repulsion at the origin 
of the density of real eigenvalues \cite{DN}.
Analytically this is seen as eq. (\ref{rho1}) matches in the Hermitian limit 
$\alpha\to 0$ with the corresponding density \cite{DN}. 

In the strong non-Hermitian limit we rescale $\sqrt{N}z=\xi$ due 
to the change of mean level spacing. Formally it can also be obtained from 
eq. (\ref{rho1}) by taking $\alpha\to\infty$. The density 
\beq
\rho^{(2)}_S(\xi) \ =\  \frac{1}{\pi(1-\tau^2)}
\left\{1\ -\ \exp\left[-\frac{1}{1-\tau^2}|\xi-i\mu|^2\right]\right\} 
\label{rho1strong}
\eeq
displayed 
in Fig. \ref{rhow+s} (right) 
shows an additional level 
repulsion at zero compared to the entirely flat density of Ginibre 
\cite{Gin} at $N_f=0$. The transition from real ($\alpha=0$) to strong 
non-Hermitian ($\alpha=\infty$) correlations resembles the results    
\cite{MPW} from lattice simulations of QCD in 
four dimensions with chemical potential.

\end{document}